\documentclass{aa}
\usepackage{epsfig,graphicx}
\usepackage{txfonts}

\usepackage{natbib}

\newcommand{\Alf}{Alfv\'{e}n }

\newcommand{\msun}{\,\mbox{$\mbox{M}_{\odot}$}}
\newcommand{\rsun}{\,\mbox{$\mbox{R}_{\odot}$}}

\newcommand{\pderiv}[2]{\mbox{${{\displaystyle\partial#1}\over {\displaystyle\partial#2}}$}}
\newcommand{\bvec}[1]{\mbox{\boldmath${#1}$}}
\begin{document}
%
   \title{Magnetic braking in young late-type stars:}
      \subtitle{the effect of polar spots}
    \titlerunning{Polar spots and magnetic braking}

   \author{A. Aib\'eo
                \inst{1,2}
                \and
                J.M. Ferreira
          \inst{1,3}
          \and
          J.J.G. Lima \inst{1,4}
          }
   \offprints{A. Aib\'eo, \email{aaibeo@demgi.estv.ipv.pt}}
   \institute{Centro de Astrof\'{\i}sica, Universidade do Porto,
         Rua das Estrelas, 4150-762 Porto, Portugal\\
         \and
         Escola Superior de Tecnologia de Viseu, Viseu, Portugal\\
         \and
   Universidade dos A\c{c}ores,
              Angra do Hero\'{\i}smo, A\c{c}ores, Portugal\\
         \and
             Departamento de Matem\'atica Aplicada, Faculdade de Ci\^encias,
         Universidade do Porto, Portugal\\
                         }

   \date{Accepted 26/05/2007; }


  \abstract
{The existence of rapidly rotating cool stars in young clusters
implies a reduction of angular momentum loss rate for a certain
period of the star's early life. Recently, the concentration of
magnetic flux near the poles of these stars has
been proposed as an alternative mechanism to dynamo saturation in order to
explain the saturation of angular momentum loss.}
{In this work we study the effect of magnetic surface flux
distribution on the coronal field topology and angular momentum
loss rate. We investigate if magnetic flux concentration towards
the pole is a reasonable alternative to dynamo saturation.}
{We construct a 1D wind model and also apply a 2-D self-similar
analytical model, to evaluate how the surface field distribution
affects the angular momentum loss of the rotating star.}
{From the 1D model we find that, in a magnetically dominated low
corona, the concentrated polar surface field rapidly expands to
regions of low magnetic pressure resulting in a coronal field with
small latitudinal variation. We also find that the angular
momentum loss rate due to a uniform field or a concentrated field
with equal total magnetic flux is very similar. From the 2D wind
model we show that there are several relevant factors to take into
account when studying the angular momentum loss from a star. In
particular, we show that the inclusion of force balance across the
field in a wind model is fundamental if realistic conclusions are
to be drawn from the effect of non-uniform surface field
distribution on magnetic braking. This model predicts that a
magnetic field concentrated at high latitudes leads to larger
Alfv\'en radii and larger braking rates than a smoother field
distribution.}
{From the results obtained, we argue that the magnetic surface field
 distribution towards the pole does not directly limit the braking efficiency of the wind.}

   \keywords{stars: late-type - stars: magnetic fields - stars: winds, starspots}

   \maketitle
%

\section{Introduction}
\label{introduction}

Rapidly rotating cool stars have a surface magnetic flux
distribution significantly different from that observed on the
Sun. Doppler imaging, and more recently, Zeeman Doppler imaging,
gives clear evidence for spots and significant surface magnetic
flux at high latitudes (e.g. \citeauthor{Donatietal99}, 1999). In
many cases, a large spot, or cluster of spots, is present at the
pole \citep{VogtPenrod83, Strassmeier02}. But clearly, spots and
magnetic fields are also present at low latitudes (e.g.
\citeauthor{Barnesetal98}, 1998). Knowledge of the large-scale
coronal magnetic topology can in principle be obtained by
extrapolating magnetic fields from the boundary data (e.g.
\citeauthor{Hussainetal01}, 2001). However, the absence of
information from a fraction of the stellar surface and the weak
correlation between spots and magnetic signatures makes us look on
the results of this promising technique with some caution.

The presence of spots and surface magnetic flux at high latitudes
has important consequences for several phenomena occurring in the stellar
corona. It can cause non-solar type
 phenomena like flares and X-ray emission at high latitudes \citep{SchmittFavata99} and
 slingshot prominences \citep{CameronRobinson89}. Here, we  concentrate our attention on
 its implications for stellar spindown.

Wind magnetic braking is based on the principle that when gas
emitted from a star is kept co-rotating with the star by magnetic
torque, it transports significantly more angular momentum outwards
than gas that conserves its angular momentum as it moves outwards
\citep{Schatzman62}. In axisymmetric winds the angular momentum
loss rate is equivalent to that carried by the gas kept in
co-rotation with the star out to the the Alfv\'en surface, where
the poloidal wind velocity equals the poloidal Alfv\'en velocity.
The presence of rapidly rotating late-type stars in young clusters
implies that there must be some limitation to the efficiency of
magnetic braking during the pre-main sequence phase (e.g.
\citeauthor{BarnesSofia96}, 1996). One possibility is that beyond
some rotation rate the magnetic field strength no longer increases
with increasing rotation rate, i.e., the dynamo saturates
\citep{MacgregorBrenner91}. But several alternatives to dynamo
saturation have been suggested. These include, the increase of the
closed field region with rotation rate \citep{MestelSpruit87}, and
the concentration of magnetic flux near the poles of rapidly
rotating stars \citep{Solankietal97, Buzasi97} can lead to a
saturation in the angular momentum loss rate. The effect of the
closed field region, or dead zone, on the rate of braking has been
studied in some detail for both single and binary stars (e.g.
\citeauthor{Lietal94}, 1994).

The idea that the concentration of magnetic flux near the poles
can mimic dynamo saturation is based on basic principles. The
Alfv\'en radius is large near the poles where the field is strong,
but smaller near the equator where the field is weaker. As wind
braking is mainly from the contributions of low and intermediate
latitudes, the overall effect is a reduction in the braking when
compared with an homogeneous surface field distribution with the
same total magnetic flux. In recent studies, the classical
\citet{WeberDavis67} wind model is explored in detail and extended
to study the effect of non-uniform surface magnetic field
distribution on the angular momentum loss rate \citep{Holzwarth05,
HolzwarthJardine05}. All these models assume that the coronal
radial field has a latitudinal distribution similar to the surface
radial field, i.e., the force balance across the field is
neglected.

In the particular case of the Sun, \textsc{Ulysses} observations
have shown that there is no significant gradient in latitude in
the radial component of the interplanetary magnetic field
\citep{Baloghetal95,SmithBalogh95}. These observations have been
explained as a result of the low plasma beta of the solar corona
so that the latitudinal and longitudinal gradients in radial
fields relax quickly, creating an essentially uniform field by
5-10 \rsun \citep{SuessSmith96}.

In the present work we consider whether polar magnetic flux
concentration towards the pole is a valid alternative to dynamo
saturation as an explanation for angular momentum loss saturation.
 The aim of this paper is also to demonstrate that the inclusion of force balance across the
 field in a wind model is fundamental if realistic conclusions are to be drawn from the effect
 of non-uniform surface field distribution on magnetic braking. Section~\ref{low_field} presents
  a 1D Weber \& Davis type wind model
 along a totally opened potential magnetic field resulting from a surface
 flux distribution concentrated towards the pole. In Sect.~\ref{wind_model} the 2D wind model
 of \citet{LPT01} is used to study the effect of the variation of surface flux with latitude on
the rate of magnetic braking. The implications and limitations of
the results obtained are discussed
 in Sect.~\ref{discussion} and the conclusions presented in Sect.~\ref{conclusions}.

\section{Application of a 1D wind model}
\label{low_field}

Present models of the inhibition of the angular momentum loss by
polar concentration of the magnetic field are qualitative in
nature and neglect force balance across the field
\citep{Solankietal97, Buzasi97, Holzwarth05}. They are based on
the assumption that the concentration of magnetic field at high
latitudes generates a large Alfv\'en radius near the pole and a
small Alfv\'en radius near the equator leading to a smaller
effective Alfv\'en radius and therefore a reduced braking
efficiency.

Ignoring force balance across the field and the indirect effect of
the magnetic field on the wind dynamics are too severe
restrictions present in these qualitative models. This makes us
doubt the conclusion that field concentration at high latitudes
significantly reduces the effective Alfv\'en radius and angular
momentum loss rate. Therefore, we present a simple wind model akin
to these qualitative models
\citep{WeberDavis67,Sakurai85,Holzwarth05} for a surface field
concentrated towards high latitudes but where force balance across
field lines is partially taken into account. In modelling the
solar corona and solar wind it is common to apply the potential
field source surface model of the coronal magnetic field
\citep{AltschulerNewkirk69, SuessSmith96}. In this model, the
magnetic field is assumed to be potential between the surface and
a spherical outer surface where the field is required to become
radial. Although this model ignores volume and surface currents,
it would be adequate to show that the latitudinal gradients of the
radial field are smoothed out at relatively short distances from
the surface (e.g. \citeauthor{Suess77}, 1977;
\citeauthor{Rileyetal06}, 2006). However, this model would be
inadequate to determine the field configuration outside the source
surface and useless if one assumes the field to be totally open.
Therefore, we use a different model to determine the coronal
magnetic field.

We construct a totally open magnetic field configuration by
considering the magnetic field to be the dominant force in the low
corona so that it is potential everywhere except at an equatorial
current sheet. We then solve the wind equations along the field
lines and compare our solutions with those that result from a
uniform surface field.

We start by considering an axisymmetric poloidal magnetic field in
an atmosphere with negligible mass and pressure. To allow for the
effect of the wind without directly solving the complicated set of
equations, a certain amount of magnetic flux is taken as open. We
then make use of a family of analytical solutions to construct
axisymmetric, partially open magnetic fields that are potential
everywhere except on a force-free equatorial current sheet
\citep{Low86, LepeltierAly96}. The magnetic field can be expressed
in spherical coordinates in terms of the stream function $A$:
\begin{equation}
\bvec{B}= \frac{1}{r\sin\theta} \left(\frac{1}{r} \pderiv{A}{\theta},-\pderiv{A}{r},0\right),
\end{equation}
so that Maxwell's equation $\bvec{\nabla . B}=0$ is satisfied. In this way, magnetic field
lines are represented by contours of constant values of the stream function, $A$.
The stream function is given by a linear combination of basic functions $Z_n$
\begin{equation}
A= \sum_{n} \gamma_n Z_n,
\end{equation}
where $n$ takes odd values, $\gamma_n$ are constant coefficients
and the analytical functions $Z_n$ are developed from the oblate
spheroidal harmonics and classified according to the harmonic
order $n$. There is a free parameter, $a$, representing the radial
distance beyond which the magnetic field is completely open.
Because these functions are not orthogonal, it is complicated to
determine the coefficients $\gamma_n$ for a prescribed boundary
condition. Nevertheless, we can combine different functions with
suitable coefficients to obtain a field with the desired
properties. The procedure to determine the different $Z_n$ is
described in Low (1986) where $Z_1$ an $Z_3$ are explicitly given.
The functions $Z_5$ and $Z_7$ are given in Appendix
\ref{stream_function}. We consider the case
$A_{dip}=\gamma_1^{dip}Z_1$, representing a dipole-like field and
$A_{polar}=\gamma_1 Z_1+\gamma_3 Z_3+\gamma_5 Z_5+\gamma_7 Z_7$,
representing a magnetic field concentrated at the poles. The
concentrated field distribution is not intended to be a realistic
field distribution but merely an extreme case of field
concentration towards high latitudes used for illustrative
purposes.

We consider fully open fields in accordance with the picture
presented in previous works \citep{Solankietal97, Buzasi97,
HolzwarthJardine05}. Thus, we set $a=r_0$ (where $r_0$ represents
the stellar radius) and use the values of $\gamma_n$ given in
Table~\ref{constants}, while the value of $\gamma_1^{dip}$ is
determined by the condition that the total magnetic flux is the
same in both cases.
\begin{table}
\begin{center}
\caption{Coefficients $\gamma_n$ for a fully open magnetic field} {\footnotesize
\begin{tabular} {@{}cc@{}}
\hline
{}&  Fully Open  \\
{} &{} \\
$ \gamma_1$ & $ 1.0$ \\
$\gamma_3 $ & $ -3.2\times 10^{-2} $ \\
$ \gamma_5$ & $ 1.4 \times 10^{-3}$ \\
$ \gamma_7$ & $ 1.0 \times 10^{-4} $ \\
\hline
\end{tabular}
\label{constants} }
\end{center}
\end{table}

\begin{figure}
   \centering
   \includegraphics[width=235 pt]{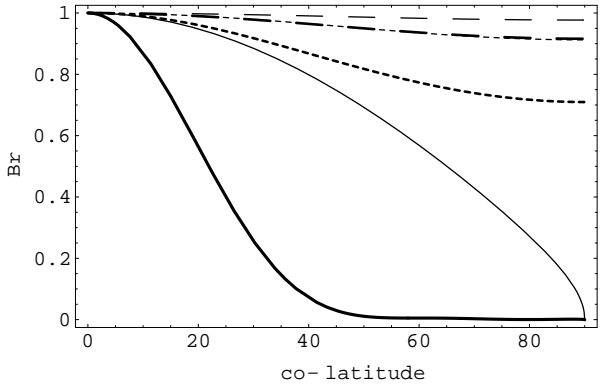}
      \caption{Fully open fields. The radial field latitudinal distribution: at the surface (full line), at $r=4r_{0}$ (short dashed) and at $r=8r_{0}$ (long dashed). The thick lines represent the  poleward concentrated field and the thin lines the dipole-like field. For a clearer representation all the radial fields are set to unity at the pole.}
      \label{radial1}
\end{figure}

\begin{figure}
   \centering
   \includegraphics[width=0.65\linewidth]{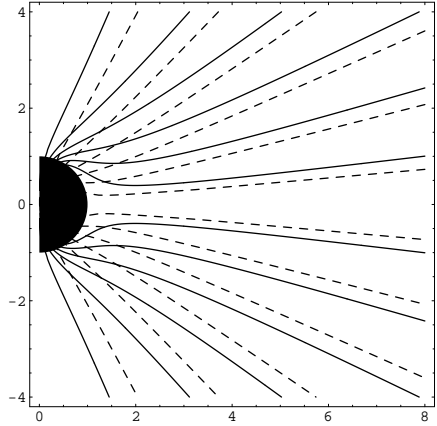}
      \caption{Fully open fields. The lines of force of the dipole-like surface flux distribution (dashed line) and the flux distribution concentrated towards the pole (full line). For the later case notice the field lines bending towards the equator.}
      \label{field_topology1}
\end{figure}

Figure~\ref{radial1} represents the radial field strength as a
function of  co-latitude for different radial distances from the
stellar center for the two cases considered. The latitudinal
profiles of the two radial fields are very different from each
other near the surface of the star. However, at intermediate
(r=4$r_{0}$) and at large distances from the surface (r=8$r_{0}$)
they are  similar to each other and to the split monopolar field.
In Fig.~\ref{field_topology1} the coronal field topology for both
cases is represented. This figure clearly shows that for surface
fields concentrated near the poles, the field lines that emerge at
high latitudes are pushed towards low latitudes in the low corona,
resulting in a field almost independent of latitude further out in
the corona. It is straightforward to show, using the same model
for $a>r_0$, that the same happens for partially open fields.

These results demonstrate that very different surface flux
distributions give rise to similar coronal fields. Physically,
this is simply a consequence of the very large magnetic pressure
difference between high and low latitudes (this is clearly
illustrated in Parker's monograph, \citet{Parker79}). Therefore, a
strong field near the pole and a weak field near the equator at
the stellar surface do not imply that the same is true in the
corona. We note that, in essence, this argument is identical to
the one used to explain \textsc{Ulysses} observations concerning
the lack of latitudinal gradients in the radial component of the
solar wind magnetic field \citep{SuessSmith96}. Having obtained
the poloidal field configuration, we now determine the polytropic
wind solution along the field. We consider a star of one solar
mass and radius, rotating rigidly and characterized by a corona
with uniform base temperature, $T=2.1\times 10^6 \mbox{K}$, and
uniform base density, $\rho=10^{-13}~\mbox{$kg/m^3$}$. We also
assume a mean atomic weight of $\tilde{\mu}=0.6$ and a polytropic
index of $\gamma=1.15$. Here we follow closely the approach
described in \citet{Sakurai85}. In brief, we solve the Bernoulli
equation presented in the Appendix~\ref{A_wind_model} for
different stellar rotation periods, ranging from 1 to 30 days, and
for different total surface magnetic fluxes, corresponding to
uniform fields ranging from 1 to 100 G. However, we do not assume
the area of each flux tube to increase as $r^2$, but we impose the
area variation to be that of the solution of the field. In
particular, we consider two flux tubes with contrasting
properties. One has its roots near the pole at latitude
$89^{\circ}$ where the field near the surface expands very
rapidly, and the other has its roots close to the equator at
latitude $1^{\circ}$ where the field near the surface first
contracts and then expands very slowly. Both flux tubes expand
$\propto r^2$ far from the surface. The latitude of the flux tube
is assumed the same and constant for all the flux tubes. We then
compare the solutions obtained with those along a flux tube with
an area variation $\propto r^2$. The solutions are compared
imposing equal flux tube area at the Alfv\'en radius as the
angular momentum loss rate is determined
 from the Alfv\'en radius, velocity and density (Eq.~\ref{aml}).
Density, velocity and radius at the Alfv\'enic point are denoted
by $\rho_*$, $v_*$ and $R_*$, respectively. We find that: i) the
Alfv\'en radius changes very little ($\le 2\%$), being larger for
rapidly expanding loops; ii) the base velocity increases
considerably as one goes from slowly expanding to rapidly
expanding flux tubes (over two order of magnitude); iii) the mass
loss rate at the Alf\'en radius changes little from flux tube to
flux tube; iv) the slow magnetosonic point decreases significantly
from slowly expanding to rapidly expanding loops (up to $25\%$);
v) the differences in angular momentum carried by the wind and
magnetic torques ($\propto \rho_* V_* R_*^4$) by the different
flux tubes of equal area at $R_*$ are negligible ($\le 2\%$).
These results determine that, in disagreement with previous
results, there is no significant difference in the angular
momentum loss when the surface field changes from being uniform to
being concentrated towards the poles. We note that centrifugal
acceleration on rapidly rotating stars is inefficient at high
latitudes and generates a non-spherical Alfv\'en surface, but
clearly, this is independent of the surface field distribution.

In the simple model presented here we observe the expansion of the polar
 field to low latitudes. However, this expansion can be limited  by a high
 equatorial gas pressure and the Lorentz force due
to the azimuthal magnetic field. Therefore, these particular
results are only valid provided the corona near the stellar
surface is a low beta plasma and the effect of the azimuthal
magnetic field can be neglected, which is likely to be the case in
magnetized rapidly-rotating stars. The parameters considered range
from those characteristic of stars termed as slow magnetic
rotators (SMR) to those characteristic of fast magnetic rotators
(FMR), as classified by \citet{BelcherMacGregor76}. The more
rapidly rotating and magnetic stars considered in the present
analysis are FMR, for which there is a poleward collimation of the
field lines \citep{HeyvaertsNorman89} that is neglected in these
simple models. But, as will be argued in detail in
Sect.~\ref{discussion}, this is largely independent of what is
under scrutiny here - the effect of the surface magnetic flux
distribution.

To complete the model just presented, two different and
complementary approaches could be pursued. In one approach,
numerical simulations are employed (e.g. \citeauthor{Sakurai85},
1985; \citeauthor{KeppensGoedbloed99}, 1999) while in the other,
analytical or semi-analytical solutions of the equations are
obtained under some simplifying assumptions. Several self similar
non-polytropic models have been developed based on the assumption
of a non-linear separation of variables (e.g.
\citeauthor{TsinganosTrussoni91:a} 1991;
\citeauthor{SautyTsinganos94}, 1994). Here, we adopt the
analytical 2-D model of \citet{LPT01} as it is the only available
analytical model that allows us to vary the magnetic surface flux
distribution and determine how it affects the wind dynamics and
the angular momentum loss rate. As in the 1-D model of Holzwarth
(\citeyear{Holzwarth05}), the poloidal field and velocity are
purely radial, but, in this case, force balance across the field
is obeyed in a self-consistent way. Although these analytical
models are more rigorous than the qualitative model just
presented, they require several simplifying assumptions that may
be physically unrealistic or undesirable. Therefore, it seems
necessary to study how such analytical models compare with
qualitative models which are more commonly applied to address the
problem of magnetic braking.

\section{Application of a 2D wind model}
\label{wind_model}

\subsection{A self-similar MHD wind model}

 In order to construct a model for an axisymmetric wind
emanating from a central rotating star, \citet{LPT01} have assumed
$\theta$-self similarity and deduced a solution of the system of
ideal MHD equations. The solution is found by a non-linear
separation of variables, keeping the treatment as general as
possible ({\it e.g.} not assuming, {\it a priori}, any prescribed
variation with latitude of the velocity, density or
magnetic field).

The model uses spherical coordinates $\ [\emph{r}, \theta, \phi]$
and assumes a simple geometry with zero meridional components of
the velocity and magnetic fields in order to find a treatable form
for the fundamental solutions an therefore calculate a solution.
The outflow dynamics are described by the following set of
equations of distance and co-latitude for the radial velocity,
azimuthal velocity, radial magnetic field, azimuthal magnetic
field, density, pressure and stellar angular velocity:
\begin{eqnarray}
    V_r(R,\theta) & = & V_0 Y\sqrt{\frac{1+\mu \sin^{2\epsilon}\theta}{1+\delta \sin^{2 \epsilon}\theta}}\label{LPT_model_begin}\\
    V_{\phi}(R,\theta) & = & \lambda V_{0} \left ( \frac{Y_{*}-Y^2}{1-M_A^2} \right) \frac{R \sin^{\epsilon} \theta}{\sqrt{1+\delta \sin^{2\epsilon}\theta}}\\
   \label{V_fi}
    B_r(R,\theta) & = & \frac{B_0}{R^2}\sqrt{1+\mu \sin^{2\epsilon}\theta}\label{B_r}\\
     B_{\phi}(R,\theta) & = & \lambda B_0 \left (\frac{R_{*}^2/R^2-1}{1-M_A^2}\right) R \sin^{\epsilon} \theta\\
    \rho(R,\theta) & = & \frac{\rho_0}{Y R^2}\left(1+\delta
    \sin^{2\epsilon}\theta\right)\\
        \label{LPT_model_mean}
    P(R,\theta) & = & \frac{1}{2}\rho_0 V_0^2\left(Q_0+Q_1 \sin^{2\epsilon}\theta
    \right)\\
    \label{LPT_model_end}
    \Omega(\theta) & = & \frac{\lambda V_{0}Y_{*}}{R_0} \frac{\sin^{\epsilon-1}\theta}{\sqrt{1+\delta \sin^{2\epsilon}\theta}}
    \label{LPT_model_omega}
\end{eqnarray}

where $\rho_0$, $B_0$, $V_0$ represent the density, radial
magnetic field and radial velocity at the polar base of the wind,
respectively, and  $M_A$, $Q_0$ and $Q_1$ are functions of $R$ and
$Y$, itself a function of $R$. The function $M_A$ is the
Alfv\'{e}n number defined by the ratio of the poloidal velocity to
the Alfv\'{e}n velocity ($M_A^2=4\pi\rho V_r^2/B_r^2$). From this
definition and from Eqs. \ref{LPT_model_begin} and \ref{B_r} it
results that the Alfv\'{e}n iso-surfaces are spherical. The
calculation of the function expressing the radial dependence of
the radial velocity, $Y(R)$, is made from the combination of the
radial and latitudinal components of the momentum equation, which
results in one first order non-linear differential equation. Such
an equation, which combines force balance along and across the
fieldlines, shows two points where both the numerator and
denominator vanish simultaneously. These singular points are
related to the non-linearity of the steady-state system of
equations. The Alfv\'{e}n point, where $M_A=1$, is a star-type
singular point and it is indicated by $(R_*,Y_*)$. All solutions
can pass through it. The second singular point is the fast
magnetosonic point and it is an X-type point allowing only two
solutions to cross it. Due to the self-similar nature of this
model, in which force balance is solved simultaneously along and
across the fieldlines, the results show that there are only the
above two critical points. Other self-similar wind models show the
same number of critical points (e.g.
\citeauthor{SautyTsinganos94}, 1994). A detailed discussion of the
nature of these critical points can be found in
\citeauthor{Tsinganosetal96}, (1996). The functions $Q_0$ and
$Q_1$ represent, respectively, the isotropic and latitudinal
dependent normalized components of the pressure (for further
details on their calculation see \citeauthor{LPT01}, 2001). The
temperature is calculated \emph{a posteriori} using the classical
ideal gas law.

The solutions of this model are defined by six dimensionless
parameters: $\lambda$, $\nu$, $M_A^0$, $\mu$, $\delta$,
$\epsilon$. The parameter $\lambda$ represents the ratio between
two velocities: the equatorial stellar rotation velocity and the
polar radial velocity at the Alfv\'en point: $\lambda= r_0 \Omega
(R_0, \pi/2)/ V_r(R_\ast,0)$ (for $\delta=0$). As a consequence of
the angular momentum conservation and induction equation, this
parameter can analogously be defined using the two components of
the magnetic field. The parameter $\nu$ is the ratio of the
stellar escape velocity and $V_0$. Thus, for a given value of
$\nu$, the parameter $\lambda$ and $Y_\ast$ define how fast or
slow rotator the star is. The parameter $M_A^0$ defines the
Alfv\'{e}n number at the wind base, i.e., it defines how
magnetized the star is. The remaining three parameters are related
to the latitudinal distribution of the different physical
quantities.

The parameter $\delta$ evaluates the density anisotropy between
the equator and the pole (cf. Eq.~\ref{LPT_model_mean}). In a
similar way, the parameter $\mu$ determines the radial magnetic
field anisotropy
 between the equator and the pole (cf. Eq.~\ref{B_r}). Finally, $\epsilon$ controls the shape of the
latitudinal distribution of the magnetic field, density, and
velocity where high values lead to steep variations and low values
lead to smoother variations.
 These three anisotropy parameters yield the flexibility of generating solutions corresponding to
  different latitudinal dependences and enable us to model stellar outflows
showing distinct latitudinal distributions of magnetic flux.

\begin{figure}
\centering
   \includegraphics[width=250 pt]{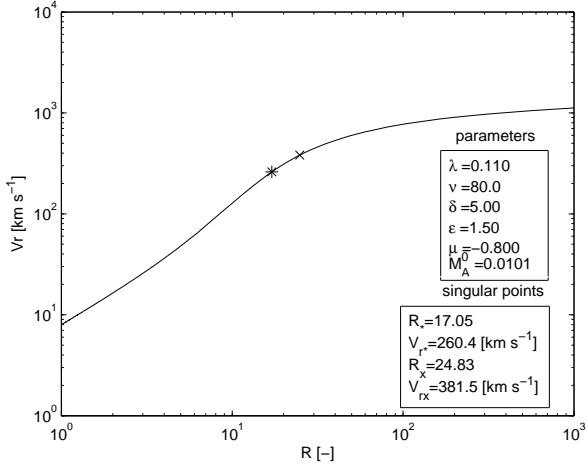}\\
  \caption{\small{A typical wind solution for a highly magnetized object in the \citet{LPT01} model, featuring two distinct singular points: the Alfv\'{e}n point and a fast magnetosonic point.}}\label{wind_example_neg}
\end{figure}

\subsection{Analysis of a typical wind solution}
\label{wind_solution}

From a purely theoretical point of view, if the concentration of
open magnetic flux reduces the efficiency of magnetic braking,
then it will do so for both fast and slow rotators. Both
observations and theoretical modelling indicate that a substantial
amount of magnetic flux concentrates at high latitudes for solar
type-stars of periods below a few days ($\sim 5$ day;
\citeauthor{Strassmeier05} 2005, \citeauthor{SchuesslerSolanki92},
1992 and \citeauthor{SchrijverTitle01} 2001). We then consider a
young, moderately rapidly-rotating solar type star with
$r_{0}=1\rsun$, $M=1\msun$, with a 4 day period and with a coronal
density $\rho_0=10\rho_\odot=1.6 \times 10^{-12}$
\mbox{kg/m$^{-3}$}.
 The parameters adopted are: ${M_A}^0=0.01$, $\nu=80$,
$\lambda=0.11$, yielding $V_0=7.72\, \mbox{km/s}$,
$V_1=0.85\,\mbox{km/s}$, and $B_0=11\,\mbox{G}$. The parameters
describing the latitudinal profiles are $\delta=5$, $\mu=-0.8$ and
$\epsilon=1.0$ that are representative of a star with more
magnetic flux at high latitudes than at low latitudes.

The wind solution obtained is presented in
Fig.~\ref{wind_example_neg} and in order to understand what forces
are relevant in accelerating the wind and in maintaining the
equilibrium in the latitudinal direction, we present a detailed
study of this solution.

The three components of the equation of motion, under the
assumptions of axisymmetry and zero theta component's are
explicitly written in Eqs.~\ref{momentum-r} to~\ref{momentum-fi}
where we have labelled the different terms with roman numerals for
easier identification.

\begin{eqnarray}
   \label{momentum-r}
   \rho V_{r}\frac{\partial V_{r}}{\partial r}- \rho \frac{V_{\phi}^2}{r}+\frac{\partial p}{\partial r}+\frac{B_{\phi}^2}{4 \pi r}+\frac{B_{\phi}}{4 \pi}\frac{\partial B_{\phi}}{\partial r} + \frac{\rho G M}{r^2} &=& 0\,\,\,\,\,\,\,\,\,\,\,\, \\
 \,\,\,\,\,\,\, I\,\,\,\,\,\,\,\,\,\, \,\,\,\,\,  II\,\,\,\,\, \,\,\,\,\,   III\,\,\,\,\,\,\, IV   \,\,\,\,\,\,\,\,\,\,\,\,\,\,\,\,  V\,\,\,\,\,\,\,\,\,\,\,\,\,\,\,\,\, VI \,\,\,\,\,\,\, & & \nonumber \\
   \nonumber\\
  \label{momentum-teta}
  \rho V_{\phi}^2\frac{\cos {\theta}}{\sin {\theta}}-\frac{\partial p}{\partial \theta}-\frac{B_{r}}{4 \pi}\frac{\partial B_{r}}{\partial \theta}- \frac{B_{\phi}^2}{4 \pi}\frac{\cos {\theta}}{\sin {\theta}} -\frac{B_{\phi}}{4 \pi}\frac{\partial B_{\phi}}{\partial \theta}&=& 0\,\,\,\,\,\,\,\,\,\\
    \,\,\,VII\,\,\,\,\, \,\,\,\, VIII\,\,\,\,\,\,\,\, \,\,\,\,\,   IX\,\,\,\,\,\,\,\,\,\,\,\,\,\,\,\,\, X\,\,\,\,\,\,\,\,\,\,\,\,\,\,\,\, \,\,\,  XI \,\,\,\,\,\,\,\,\,\,\,& & \nonumber \\
   \nonumber \\
    \label{momentum-fi}
   \frac{\rho V_{r}}{r}\frac{\partial (r
V_{\phi})}{\partial r}-\frac{B_{r}}{4 \pi r}\frac{\partial(r
B_{\phi})}{\partial r}&=& 0
\end{eqnarray}

\begin{figure*}
\centering
   \includegraphics[width=480 pt, height=210 pt]{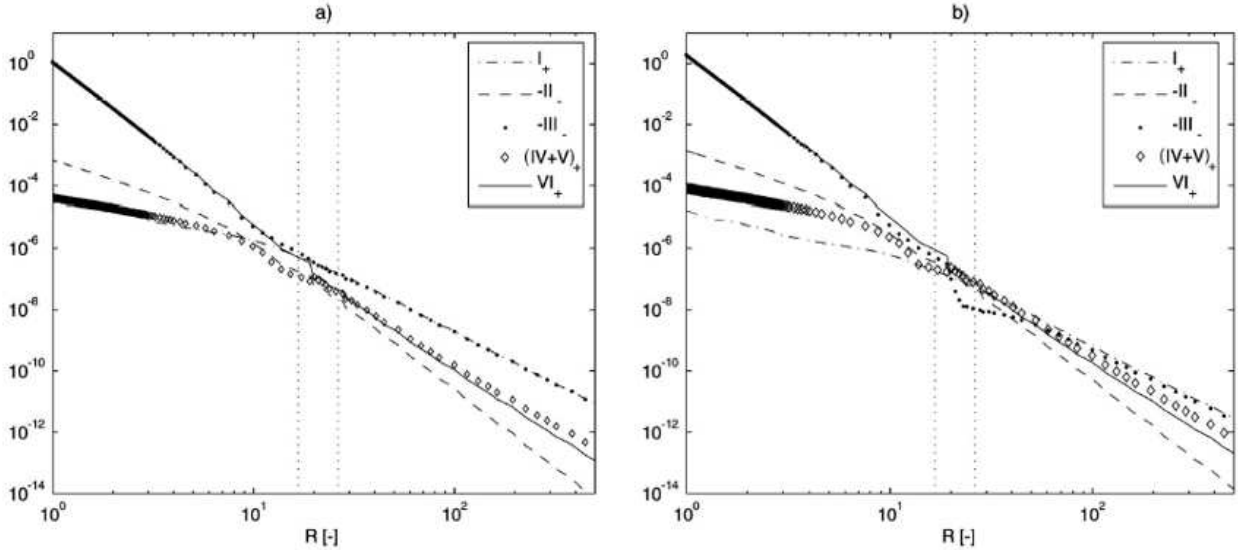}\\
  \caption{Absolute values of the different terms of the radial component
   of the force balance equation as a function of radial distance. Positive terms
   are labelled by (+) and negative terms by (-). In panel a) at $45^\circ$ of latitude. In
   panel b) at the equator.}\label{r_panel}
\end{figure*}

In Fig.~\ref{r_panel} we can observe that close to the surface the
dominant forces in the radial direction are the gas pressure
gradient (III) and gravity (VI), which balance each other in
almost hydrostatic equilibrium. Further out, at high and
intermediate latitudes (panel a) of Fig.~\ref{r_panel}), it is the
pressure gradient that accelerates the wind with a small
contribution from the other forces. However, near the equator both
 the centrifugal force (II) and the
Lorentz force (IV+V) due to the $\phi$-component of the field are
important in accelerating the wind near and beyond the singular
points.
 This is in accordance with what one would expect for a star with the properties considered here.

\begin{figure}
\centering
   \includegraphics[width=250 pt]{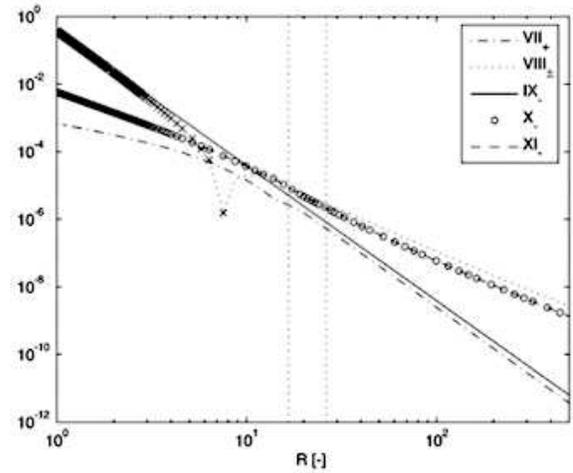}\\
  \caption{\small{Absolute values of the different terms of the $\theta$-component of the force balance equation as a function of radial  distance at the pole. Positive terms are labeled by (+), negative terms are labeled by (-) and terms that change sign by ($\pm$) together with a cross along the part where they are negative.}}\label{teta_90}
\end{figure}

It is also instructive to analyze the force balance in the
$\theta$-direction. Figure~\ref{teta_90} shows that below the
critical points
 the magnetic pressure
gradient due to the radial magnetic field (IX) is balanced by the
gas pressure gradient (VIII) and, to a lesser extent, by the
Lorentz force due to the $\phi$-component of the field (X+XI). The
gas pressure gradient decreases and changes sign close to the \Alf
point, as indicated by the crosses on the dotted curve. In this
region the force opposing the radial magnetic field pressure
gradient is the Lorentz force resulting from the $\phi$-component
of the magnetic field (X + XI). Therefore, the expansion of the
poloidal field from the high latitudes towards the low altitudes
is prevented by both the gas pressure gradient and the toroidal
component of the magnetic field. This is possible because the
plasma $\beta$ ($\beta=p/(B^2/8\pi)$) at the surface is relatively
high, varying from $0.16$ at the pole to $4.7$ at the equator.
Otherwise, in a low $\beta$ corona, the field lines would bend
towards regions of lower  magnetic pressure
 so that different magnetic surface distributions would generate
similar coronal topologies, as discussed in Sect.~\ref{low_field}.
Beyond the critical points, the Lorentz force that tends to
collimate the field lines towards the poles (IX) is balanced by
the gas pressure gradient (VII), so that the field lines remain
radial in the poloidal plane. For a significantly more magnetic
and more rapidly rotating star than the one considered here, there
is no force capable of balancing the Lorentz force and the model
cannot generate physically acceptable solutions. In other words,
the assumptions of no meridional components of magnetic field and
velocity and of a full open magnetic field is incompatible with a
magnetically dominated low corona.

\begin{figure}
\centering
   \includegraphics[width=250 pt]{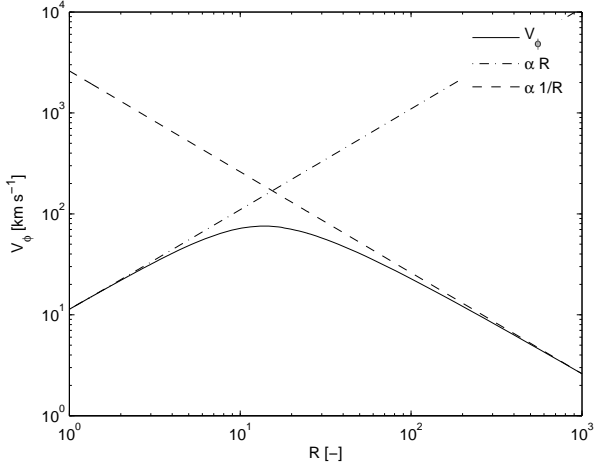}\\
  \caption{\small{Azimuthal velocity at the equator as a function of radial distance. Also represented are the profiles corresponding to co-rotation (dot-dashed line) and angular momentum conservation (dashed line)}}\label{Vfi_bounded}
\end{figure}

The $\phi$-component of the equation of motion expresses that the
change of angular momentum of the wind is equal to the magnetic
torque. As a consequence, the azimuthal velocity is close to
co-rotation near the surface due to the strong magnetic torque,
while far from the surface this torque becomes less effective and
the conservation of angular momentum implies that the azimuthal
velocity must decrease proportionally to $1/R$, as is illustrated
in Fig.~\ref{Vfi_bounded}.

\begin{figure}
   \includegraphics[width=250 pt]{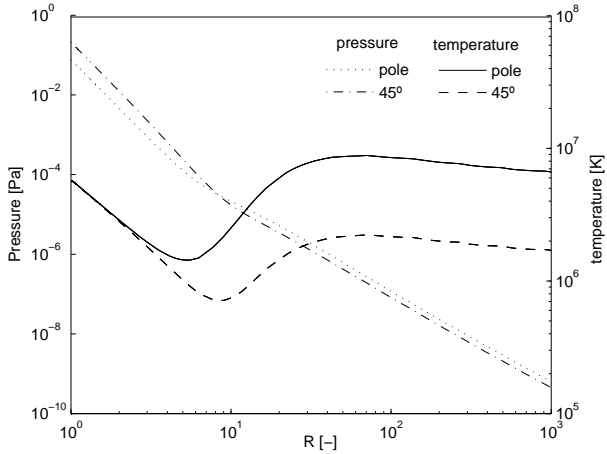}\\
  \caption{\small{Radial profiles of pressure and temperature at $45^\circ$ of latitude and pole.}}\label{p_e_T}
\end{figure}

\begin{figure}
  \includegraphics[width=250 pt]{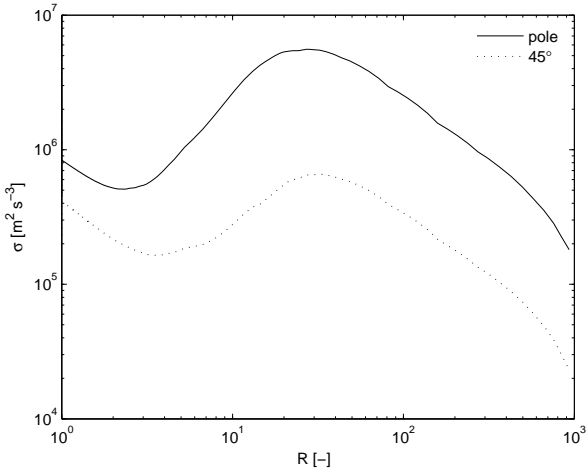}\\
  \caption{\small{Radial profile of the heating rate per unit of mass at $45^\circ$ of latitude and pole.}}\label{sigma}
\end{figure}

Figure \ref{p_e_T} represents the gas pressure and temperature at
two different latitudes as a function of radial distance. The
pressure close to the surface increases towards the equator but
has the opposite behaviour further out, i.e., as pointed out
earlier, the gas pressure gradient changes sign close to the \Alf
point. In this model, the coronal temperature increases towards
the pole as a consequence of the latitudinal behaviour of the gas
pressure and density. Having in mind that the temperature is not
imposed but obtained {\it a posteriori } from the perfect gas law,
its values of $T\approx 10^6 - 10^7 \rm{K}$ are in relatively good
agreement with what we would expect for an active solar-type star.
It is usually assumed in numerical models of solar and stellar
winds that the corona is isothermal, or, more generally, it
follows a polytropic law. In our approach it is not possible to
impose this behaviour and it is also not desirable, as stellar
coronae are not well described by such simple laws. Additionally,
it is not expected that coronal regions of large and small
magnetic field concentration have the same temperature. Instead,
the temperature profile allows us to deduce, from a consistent
solution of the energy equation, the regions where energy is
deposited or removed
\begin{equation}\label{energy_eq}
    \sigma=V_r\frac{\partial}{\partial R}\left(\frac{P}{(\Gamma-1)\rho} \right)+P V_r\frac{\partial}{\partial R}\left(\frac{1}{\rho}
    \right).
\end{equation}
where $\Gamma$ is the ratio of specific heats and $\sigma$
represents the net effect of all sources and sinks of energy per
unit of mass. In the particular case of the solution presented
here, the temperature varies with latitude and with radius. From
Figs.~\ref{p_e_T} and \ref{sigma} we can infer that the
temperature and heating rate are higher at high latitudes than at
low latitudes and that there must be a heating mechanism that
deposits energy at large radii, beyond the sub-Alfv\'enic region.
Although this model has no specific heating mechanism included, we
note that Alfv\'en waves dissipation is a viable mechanism for
energy deposition far from the surface, generating temperature
profiles somewhat resembling the ones presented here
\citep{Cranmer05}.

\begin{figure}
\centering
   \includegraphics[width=240 pt]{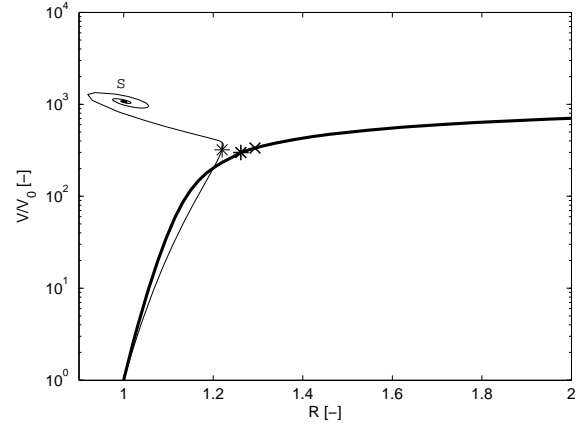}\\
  \caption{\small{Comparison between solutions for positive and negative values of the parameter $\mu$. The solid bold line represents the solution typical topology for wind type critical solutions ($\mu=-0.05$). The narrow solid line represents the topology for solutions with positive values of $\mu$ ($=0.005$), which are not wind solutions. The \textbf{*} markers stand for star-type points (the Alfv\'en points), the \textbf{x} marker represents the x-type point and the \textbf{s} marker points out the spiral-type critical point.}}\label{critical_comparison}
\end{figure}

The exploration of the parameter space
reveals some important results concerning the overall behaviour of
these wind solutions. In general, an arbitrary choice of parameters of the model
 will not generate wind solutions. In particular for the
anisotropy parameters, $\mu$ and $\delta$ have a crucial influence
in the properties of the solutions. On one hand, accelerating wind
solutions require the density to be higher at the equator than the
pole, i.e., positive values of $\delta$. On the other hand, there
are meaningful solutions only for $\mu<0$. If $\mu$ takes positive
values, the critical solution
 shows a different fast magnetosonic critical point - a spiral type point,
 preventing any possibility of building wind-type solutions. This is shown
  in Fig. \ref{critical_comparison}
 where a comparison between a typical
wind solution for a negative value of $\mu$ and a
terminated solution for a positive value of $\mu$ is presented.

The effect of the different parameters on the wind solution can
only be understood by taking into account force balance along the
lines (radial and azimuthal components of the momentum equation,
Eqs. \ref{momentum-r} and \ref{momentum-fi}) but also from the
equilibrium across the lines ($\theta$-component of the momentum
equation, Eq. \ref{momentum-teta}). As an example, if we change
the parameters in order to describe an even more magnetic and more
rapidly-rotating star, we would not be able to obtain a solution.
Physically, this results from lack of force balance across the
field. The low gas pressure can neither balance the tendency of
field lines to bend towards low latitudes at small distances from
the surface, nor the tendency of field lines to collimate at large
distances as required by the assumptions of the model.

A complete study of the behaviour of the wind solution with
the variation of the parameters is
extremely complex because the influence each parameter has on the
critical solution is often dependent on the values taken by the
other parameters. This particular study is beyond the scope of the
present work.

\subsection{Variation of angular momentum loss with the distribution of the surface field}
\label{aml_distribution}

In order to compare the angular momentum loss rate for different
 distributions of the magnetic
field at the surface of the star, we must prescribe a
fixed value of total magnetic flux, $F_0$. Therefore, the surface
magnetic field strength at the pole must be defined accordingly:

\begin{equation}\label{B_0}
    B_0=\frac{F_0}{4\pi r_0^2 \int_0^{\pi/2} \sin{\theta} \sqrt{1+\mu \sin^{2\epsilon}{\theta}}d\theta}.
\end{equation}

The mass loss rate per unit solid angle at co-latitude $\theta$ is
\begin{equation}\label{mass_flux}
    -\dot{m}(\theta) = \rho V_r r^2 =\rho_{0} V_0 r_0^2 \sqrt{(1+\mu \sin^{2\epsilon}{\theta})(1+\delta    \sin^{2\epsilon}{\theta})}.
\end{equation}
The total angular momentum loss rate is equivalent to that carried
by the gas kept in co-rotation with the star out to the Alfv\'en
surface, which can be expressed as
\begin{equation}\label{aml}
    -\dot{J}(\theta) = 4\pi \int_0^{\pi/2} \Omega(\theta) \rho_* V_* R^4_* \sin^3\theta d\theta.
\end{equation}
Or equivalently, in terms of the model parameters (Lima et al.,
2001) as
\begin{equation}
 -\dot{J}(\theta) = \lambda r_{0}^3 B_{0}^2 \int_0^{\pi/2} \sin^{\epsilon+2}{\theta} \sqrt{1+\mu \sin^{2\epsilon}{\theta}}d\theta.
\end{equation}
In this model, $\rho$, $V_r$ and $\Omega$ are functions of
$\delta$ and yield an angular momentum loss rate that is not
directly dependent on this anisotropy parameter. By changing the
anisotropy parameters $\mu$ and $\epsilon$ the Alfv\'en radius and
Alfv\'en velocity change.

\begin{figure}
\centering
   \includegraphics[width=250 pt]{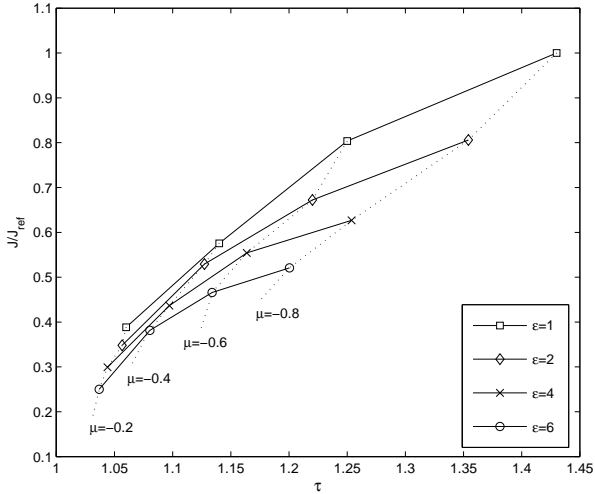}\\
  \caption{\small{Total angular momentum loss variation as a function of polar field concentration, for different sets of $\mu$ (dashed lines) and $\epsilon$  (solid lines). }}\label{aml_vs_tau}
\end{figure}

In order to evaluate the polar concentration of a given magnetic
field surface distribution, we first compare the magnetic flux at
low co-latitudes (from $30^\circ$ towards $0^\circ$) with the
overall total flux. Thus,
\begin{equation}\label{omega}
       \omega =  \frac{\int_{0}^{\pi/6} {\sqrt{1+\mu \sin^{2\epsilon}\theta}}d\theta}{\int_{0}^{\pi/2} {\sqrt{1+\mu
        \sin^{2\epsilon}\theta}}d\theta}\, .
\end{equation}

Then, we define the parameter $\tau$ that evaluates the magnetic
field concentration towards high latitudes by comparing it with a
uniform magnetic field distribution (the split-monopole case with
$\epsilon$ =1 and $\mu=0$)

\begin{equation}\label{tau}
       \tau  = \frac{\omega}{\omega_{\mbox{monopole}}}= 3 \frac{\int_{0}^{\pi/6} {\sqrt{1+\mu \sin^{2\epsilon}\theta}}d\theta}{\int_{0}^{\pi/2} {\sqrt{1+\mu
        \sin^{2\epsilon}\theta}}d\theta} \, .
\end{equation}

If $\tau>1$ the field is concentrated towards the pole while if $\tau <1$ the field is concentrated towards the equator.

As a result of the way the model is constructed, a change in the parameters $\mu$ or $\epsilon$ will change the
total amount of magnetic flux, the location of the Alfv\'en radius
 and also the rotation period of the star (see Eqs.~\ref{B_0} and \ref{LPT_model_omega}).
Therefore, an iterative procedure on the parameters $\lambda$ and
$M_A$ is performed so that the different solutions under
comparison have the same total amount of magnetic flux and
rotation period within an error of $5\%$.

By evaluating the angular momentum loss rate for different wind
solutions attained for several sets of $\mu$ and $\epsilon$, we
 are able to obtain a relation between the magnetic polar field
 concentration and the total angular momentum loss. This is shown in
Fig.~\ref{aml_vs_tau}, where solid lines represent how $-\dot{J}$
varies with $\tau$ due to the variation of $\mu$, with constant
$\epsilon$. Physically, this corresponds to determining how the
variation of the magnetic field distribution with a fixed rotation
profile affects the angular momentum loss rate. Also represented
by dashed lines is the case of how $-\dot{J}$ varies with $\tau$
due to the variation of $\epsilon$, with constant $\mu$. In this
case, both the magnetic field and rotation latitudinal profiles
vary. All solutions considered here are lied as SMR, revealing
that centrifugal forces are not predominant for the wind
acceleration mechanisms. The dynamics of the wind is, as
previously shown, independent of the forces generated by the
rotation of the central body. This is true even for the particular
case where the period is of $4$ days since other contributions,
such as the pressure gradient, are more important.

Our purpose is to study young and rapidly rotating stars,
therefore we must consider small values of $\epsilon$ and small
values
 of $\delta$, as this kind of object shows
near rigid body surface rotation with a slightly faster equator
(e.g. \citeauthor{CameronDonati02}, 2002). Also, it is only
adequate to compare stars with different surface magnetic field
distribution but with the same surface rotation latitudinal
profile. Thus, we restrict the physical application of this model
to the cases of constant $\epsilon$ and variable $\mu$. If we
analyze the solid lines  in Fig.~\ref{aml_vs_tau}, we see that the
total angular momentum loss is an increasing function of $\tau$.
Therefore, the higher the magnetic field polar concentration, the
more efficient magnetic braking is. Referring to Eq.~\ref{aml},
and noticing that, by continuity,
 $\rho_* V_* R^4_*= \rho_0 V_0 R^2_0 R^2_*$,
  changes in total angular momentum loss must arise from changes
   in the mass loss rate or changes in the Alfv\'en radius. But as the mass loss
   rate decreases with increasing $\tau$ and $|\mu|$ (cf. eq.\ref{mass_flux}), we
   can conclude that the increase in total angular momentum loss is due to an
   increase in the Alfv\'en radius that more than compensates the decrease in mass loss.

As $|\mu|$ increases, $B_0$ increases and $M_{A}^{0}$ decreases,
i.e., at the pole the star becomes more magnetic as the field
concentrates there. As
 $V_* R_*^2 \propto (M_A^0)^{-2}$ we see that either the Alfv\'en velocity, or
 the Alfv\'en radius or both must increase.
 From the different solutions present
 in Fig.~\ref{aml_vs_tau} as well as an analysis of the differential
 equation
 yielding $V_r(R)$ \textbf{(or $Y(R)$)}, we find that the wind acceleration decreases as $\mu$ increases
 so that $V_*$ decreases and the Alv\'en radius must increase. Physically, this increase
  in the Alfv\'en radius is due to a less efficient acceleration mechanism. The analysis
  and clear understanding of this result is complicated by the fact that as $\mu$ changes
  there is a direct effect on the radial magnetic field but also on the radial velocity
  latitudinal profile and, indirectly, the wind pressure and temperature. Therefore, the increase
  in angular momentum cannot be solely attributed to the increase of magnetic field
  concentration as this cannot be separated from changes in other physical
  quantities. This result is at odds with the result of Sect.~\ref{low_field}
where no significant difference in angular momentum loss as a
function of field concentration was found. This discrepancy may be
explained by the fact that the two models address totally
different scenarios characterized by different plasma $\beta$
regimes.

\section{Discussion}
\label{discussion}

Empirical or qualitative models of magnetized stellar winds have
been very successful in establishing the basic physical principles
that govern the rate of angular momentum loss associated with a
wind. In particular, the work of \citet{Mestel68} and
\citet{MestelSpruit87} investigates how the magnetic braking
efficiency varies with stellar rotation rate. It shows that it is
possible to have saturation in the angular momentum loss rate with
rotation without dynamo saturation as a result of two competing
effects. On the one hand, the magnetic torque increases with rotation
 rate due to the increase of the field strength.
 On the other hand, the fraction of the stellar surface
with open magnetic field lines contributing to braking decreases
with rotation rate. This model has also been applied to establish
that a rearrangement of the surface field from a low order to a
high order multipole at high rotation rates implies a decrease in
braking efficiency \citep{TaamSpruit89}.

There are, however, limitations to how much information can be
obtained from these qualitative models. In this work, we simply
address the question of whether a correct estimate of the
influence of the magnetic surface field concentration towards the
poles can be obtained using these empirical models or whether such
a goal requires a computation of the field that takes into account
force balance in all directions. To this end, in
Sect.~\ref{low_field} we determine the coronal magnetic field due
to a surface flux distribution concentrated around the poles and
show that this distribution is not maintained further out.
Instead, the magnetic field rapidly expands to regions of low
magnetic pressure and approaches the simple split-monopole field.
Flux tubes at different latitudes have very different expansion
rates, but we find that this has no significant effect on the
angular momentum transported outwards.

In order to investigate the properties of stellar winds for
different surface flux distributions we apply, in
Sect.~\ref{wind_model}, the analytical wind model of Lima et al.
(2001). To our knowledge, this is the first application of an
analytical quantitative 2D model to the problem of magnetic
braking in young solar- type stars. This model has very appealing
characteristics has it describes the outflow
 of a rotating star for which the
latitudinal magnetic flux distribution is the same at all radial
distances. It also has some limitations that are typical of this
kind of model. It is constructed under the assumption of
separation of variables and there is no energy equation, with the
temperature being determined from the perfect gas law. Also,
although the model allows different surface flux distribution, its
range of variation is somewhat limited. Its most restrictive
properties, however, are that the \Alf surface is spherical,
contrary to what is expected for fast rotators and has been
proposed by \citet{Solankietal97}, and the absence of meridional
components of the velocity and magnetic field. Therefore, the
conclusions obtained from using this model cannot be \textit{a
priori} considered  as general features. Yet, some of the results
are extremely relevant as they convey fundamental physical
principles. We find that the surface concentration of magnetic
flux at high latitudes can only occur in the corona for a
relatively high plasma beta, so that the gas pressure gradient can
oppose the Lorentz force that tends to bend the field lines
towards the equator. This work stresses the importance of the
latitudinal profiles of the physical quantities involved in the
description of the hydro-magnetic wind. As shown in section
\ref{wind_solution}, the radial features of the wind, such as the
acceleration mechanisms, are not only dependent on the force
balance along the lines but also on the force balance across the
lines. Furthermore, a given radial field latitudinal profile
implies, due to force balance and under the assumptions of the
model, a certain latitudinal profile of the density and velocity
that are as important as the magnetic field in determining the
angular momentum loss rate. The most important prediction from
this model is that a higher polar field concentration leads to
larger braking rates than a smoother field distribution. However,
because of the inherent coupling of this result with some of the
assumptions of the model, this cannot be viewed as general. In
spite of this, we have presented an example of an equilibrium
model, with unknown stability properties, for which magnetic
braking increases as the field concentration to the pole
increases. Remarkably, this model has the attractive property of
having a magnetic topology identical to the models of
\citeauthor{Holzwarth05} (2005), \citeauthor{HolzwarthJardine06}
(2006) and \citeauthor{Solankietal97}, (1997) but it complies with
force balance in all directions, which generates antagonistic
results.

Both models presented in this work neglect the poleward
collimation of the field lines characteristic of FMR. An
interesting question is whether this poleward collimation
increases or decreases the angular momentum loss of the star, but
we are not aware of any work specifically addressing this issue.
At high latitudes $R_*$ increases due to the effect of collimation
but the opposite happens near the equator (e.g.
\citeauthor{Sakurai85}, 1985), so that the end result depends on
which effect is dominant. In Sect.~\ref{low_field} we have
determined that at $4-8$ stellar radii from the stellar surface
the radial field becomes uniform in latitude, independently of its
surface distribution, and that in general this occurs very much
inside the Alfv\'{e}n surface. Therefore, we expect that field
collimation would occur identically for both uniform and highly
concentrated surface field distributions with no significant
differences in angular momentum loss.

Based on the results presented in Sect.~\ref{low_field} \&
\ref{wind_model}, we argue that the concentration of magnetic flux
at high latitudes does not directly contribute to limit the
braking efficiency of the wind. By comparing the results obtained
with the two different models one could in principle be able to
infer how crucial the limitations introduced are in the analytical
treatment of the 2D model to the results obtained. The extent to
which this comparison can be made is limited by the very different
assumption about the plasma $\beta$ in the two cases. We
hypothesize that the increase in angular momentum loss with
magnetic field concentration observed in Sect.~\ref{wind_model}
results from changes in the gas temperature, density and pressure
that are generated to maintain force balance across the field in
the absence of meridional components. One can speculate about what
to expect if some of the most stringent assumptions of our 2D
model are relaxed. We suggest that allowing for meridional
components, non spherical Alfv\'en surface and a more realistic
plasma $\beta$ would lead to closed magnetic field lines near the
equator. Furthermore, a surface field concentrated towards the
pole would have a weaker field at low latitudes and consequently
have more open magnetic flux than a smooth surface magnetic field.
This would again imply larger braking rates for a surface field
concentrated towards the pole than for a smooth field with equal
amounts of magnetic flux, if the wind properties are similar in
the two cases. However, the existence of large amounts of magnetic
flux at high latitudes can still lead to a reduced angular
momentum loss rate if it creates a complex field topology akin to
multi-order magnetic fields (cf. \citeauthor{TaamSpruit89}, 1989).

The work presented here can be extended by further analytical
modelling or numerical simulations. However, many aspects are
still unknown and this constrains how much one can predict about
magnetic braking. It is still not known whether polar spots are
largely unipolar, as assumed here (for opposing views see e.g.
\citeauthor{SchrijverTitle01}, 2001 and \citeauthor{McIvoretal03},
2003). In addition, we cannot rule out the possibility that
stellar winds are accelerated by dissipation of magnetic waves as
well as thermal and centrifugal forces. The nature of the slow
solar wind is still largely unknown and it remains an open
question whether it is admissible to ignore the contribution of
the slow wind to the angular
 momentum evolution of active late-type stars.

\section{Conclusions}
\label{conclusions}

In the present work we investigate how the magnetic surface field
distribution affects the coronal magnetic field and the rate of
angular momentum removal by the stellar wind.

There are three important results from our work: First, we have
shown that very different surface flux distributions yield similar
coronal fields in a low $\beta$ coronal plasma as well as similar
wind braking rates. Second, we have demonstrated that the radial
features of the wind, such as the acceleration mechanisms and the
gas pressure distribution, are not only dependent on the force
balance along the field lines, but also on the force balance
across the field lines. Finally, in the wind model of
\citep{LPT01} a higher polar field concentration leads to larger
braking rates than a smoother field distribution. However, we
cannot rule out the possibility that this is a result of the
assumptions of the model and so it may not be regarded as a
general feature. This model also demonstrates that the rate of
braking is dependent on the latitudinal behaviour of several
physical quantities and not only of the surface radial field.

We conclude that the concentration of magnetic flux at high
latitudes is unlikely to directly constrain the braking efficiency
of the wind. It may, however, have a decisive importance in
determining the amount of open magnetic flux that contributes to
wind magnetic braking.

\begin{acknowledgements}
This work was partially supported by grant POCI/CFE-AST/55691/2004
approved by FCT and POCI, with funds from the European Community
program FEDER. J. M. Ferreira wishes to thank Volkmar Holzwarth
for a profitable exchange of ideas during Cool stars XIV. The
authors would like to thank the Referee for several useful
suggestions and references on the corona and solar wind.
\end{acknowledgements}

\appendix
\section{Stream functions}
\label{stream_function}

Here we give the explicit forms of $Z_n$ for $n=5$ and $n=7$:

\begin{eqnarray}
Z_5 & = & \frac{15}{14}r \left(1-v^2\right) \left(21v^4-14v^2+1\right) \nonumber \\
&\times & \left[15\left(1+u^2\right)\left(1
 + 14u^2+21u^4\right)\tan^{-1}\frac{1}{u}\right] \nonumber \\
& - & \frac{225}{16}\frac{\pi a^2}{r} \sin^2\theta
 -  \frac{1675}{32} \frac{\pi a^4}{r^2} \sin^2\theta \left(5\cos^2\theta-1\right) \nonumber \\
& - &\frac{4725}{128} \frac{\pi a^6}{r^5} \sin^2\theta \left(1-14\cos^2\theta+21\cos^4\theta \right)\nonumber \\
&+& 30a\eta
\end{eqnarray}
\begin{eqnarray}
    Z_7 & = & \frac{7}{16} r\left(1-v^2\right)\left(429v^6-495v^4+135v^2-5\right) \nonumber \\
    & \times & \left[\left(429u^6+495u^4+135u^2+5\right)\tan^{-1}\frac{1}{u} \right. \nonumber \\ &+& \left. \frac{1}{80}\left(1837u+14273u^3+27335u^5+15015u^7\right)\right] \nonumber \\
    &+& \frac{3}{4}\frac{\pi a^2}{r}\sin^2\theta+\frac{6615}{32}\frac{\pi a^4}{r^3}\sin^2\theta \left(5\cos^2\theta-1\right)\nonumber \\
    &+ & \frac{8085}{2} \frac{\pi a^6}{r^5} \sin^2\theta \left(1-14\cos^2\theta +21\cos^4\theta \right)\nonumber \\
    &+& \frac{21021}{512}\frac{\pi a^8}{r^5}\sin^2\theta \left(-5+135\cos^2\theta-495\cos^4\theta \right. \nonumber \\
    &+& \left. 429\cos^6\theta\right) +56a \eta,
\end{eqnarray}
with

\begin{equation}
u^2=-\frac{1}{2}\left(1-\frac{a^2}{r^2}\right)+\frac{1}{2}\left[\left(1-\frac{a^2}{r^2}\right)^2+\frac{4a^2}{r^2}\cos^2\theta\right]^{1/2},
\end{equation}
\begin{equation}    v^2=-\frac{1}{2}\left(\frac{a^2}{r^2}-1\right)+\frac{1}{2}\left[\left(\frac{a^2}{r^2}-1\right)^2+\frac{4a^2}{r^2}\cos^2\theta\right]^{1/2},
\end{equation}
\begin{equation}
\eta^2=-\frac{1}{2}\left(\frac{r^2}{a^2}-1\right)+\frac{1}{2}\left[\left(\frac{r^2}{a^2}-1\right)^2+\frac{4r^2}{a^2}\cos^2\theta\right]^{1/2}.
\end{equation}

\section{Wind model}
\label{A_wind_model} Here we present how the wind solution is
obtained. As we follow very closely the approach of
\cite{Sakurai85}, only a brief description is given. We assume
azimuthal symmetry and by combining the equations of mass and
magnetic flux conservation, the frozen-in condition, the
polytropic law and the momentum conservation equation in the
poloidal and azimuthal directions we arrive at a Bernoulli
integral of the equation of motion
\begin{eqnarray}
H & = & \frac{v_p^2}{2}-\frac{\Omega^2 (r_*\sin\theta)^2}{2}\left[\left(\frac{\frac{r_*}{r}-\frac{r}{r_*}}{\frac{\rho}{\rho_*}-1}\right)^2- \left(\frac{r}{r_*}\right)^2\right]\nonumber \\
&+& \frac{\gamma}{\gamma-1} \frac{p}{\rho}-\frac{GM}{r}
 =E,
\end{eqnarray}
where $v_p$ is the poloidal velocity, $\Omega$ the stellar angular velocity, $M$ the mass of the star, $G$ the gravitational constant, $E$ an integration constant and the subscript $*$ stands for the Alfv\'en point.
Upon applying the law of mass conservation,  $\rho V_p S= \rho_0 V_{p_0} S_0$, with $S$ representing the area of the flux tube and the subscript $0$ the values at the stellar surface, we obtain $H=H(r,\rho)$. At the fast and slow critical points one has the regularity conditions
\begin{equation}
\label{zero_derivative}
\frac{\partial H}{\partial \rho} = \frac{\partial H}{\partial r}=0,
\end{equation}
together with
\begin{equation}
\label{bernoulli}
H(r_s,\rho_s)=H(r_f,\rho_f)=E.
\end{equation}
Applying mass and flux conservation allows us to write the density at the Alfv\'en point as
\begin{equation}
\rho_*= \rho_0 \frac{V^2_0}{V_{A_0}^2}\, .
\end{equation}
The system of six simultaneous algebraic equations
(\ref{zero_derivative}, \ref{bernoulli}) is solved for the
unknowns $r_*$, $r_s$, $\rho_s$, $r_f$, $\rho_f$ and $v_0$. The
integration constant $E$ is not an additional unknown as it can be
written in terms of $V_0$ and $r_*$. We note that, contrary to
\citet{Sakurai85}, we solve our equations in dimensionless form
with respect to surface values and not with respect to the
Alfv\'en radius as, in general, $S\propto r^2$ does not hold.

\bibliography{bibdata}
\bibliographystyle{NBaa}

\end{document}